\documentclass[fleqn,usenatbib]{mnras}

\usepackage{newtxtext,newtxmath}
\usepackage[T1]{fontenc}
\usepackage{ae,aecompl}

\usepackage{graphicx}	% Including figure files
\usepackage{amssymb}	% Extra maths symbols

\newcommand{\unit}{1\!\!1}

\def\vel{\, \mathbf v}
\def\Bfield{\, \mathbf B}
\def\Efield{\, \mathbf E}
\def\Fpart{\, \mathbf F_{\rm part}}
\def\upart{\, \mathbf u_{\rm part}}
\def\npart{\, n_{\rm part}}
\def\Jpart{\, \mathbf J_{\rm part}}

\def\MS{\, M_{\rm s}}
\def\MA{\, M_{\rm A}}
\def\rl{\, R_{\rm l}}

\title[particle acceleration in low-$\MS$, high-$\beta$ shocks]{On the influence of supra-thermal particle acceleration on the morphology of low-Mach, high-$\beta$ shocks}

\author[A.J. van Marle]{
Allard Jan van Marle,$^{1,2}$\thanks{E-mail: allard-jan.van-vanmarle@umontpellier.fr}
\\
$^{1}$Department of Physics, UNIST, UNIST-gil 50, Ulsan, 44919, Republic of Korea \\
$^{2}$LUPM $–$ UMR5299, Universit{\'e} de Montpellier, Campus Triolet, Place Eugene Bataillon, CC 72, 34095 Montpellier C{\'e}dex 05 FRANCE\\
}

\date{Accepted XXX. Received YYY; in original form ZZZ}

\pubyear{2020}

\begin{document}
\label{firstpage}
\pagerange{\pageref{firstpage}--\pageref{lastpage}}
\maketitle

\begin{abstract}
When two galaxy clusters encounter each other, the interaction results in a collisionless shock that is characterized by a low (1-4) sonic Mach number, and a high Alfv{\'e}nic Mach number. Our goal is to determine if, and to what extent, such shocks can accelerate particles to sufficient velocities that they can contribute to the cosmic ray spectrum.
We combine two different computational methods, magnetohydrodynamics (MHD) and particle-in-cell (PIC) into a single code that allows us to take advantage of the high computational efficiency of MHD while maintaining the ability to model the behaviour of individual non-thermal particles. Using this method, we perform a series of simulations covering the expected parameter space of galaxy cluster collision shocks. 
Our results show that for shocks with a sonic Mach number below 2.25 no diffusive shock acceleration can take place because of a lack of instabilities in the magnetic field, whereas for shocks with a sonic Mach number $\geq\,3$ the acceleration is efficient and can accelerate particles to relativistic speeds.
In the regime between these two extremes, diffusive shock acceleration can occur but is relatively inefficient because of the time- and space-dependent nature of the instabilities.
For those shocks that show efficient acceleration, the instabilities in the upstream gas increase to the point where they change the nature of the shock, which, in turn, will influence the particle injection process.
\end{abstract}


%%%%%%%%%%%%%%%%%%%%%%%%%%%%%%%%%%%%%%%%%%%%%%%%%%

%%%%%%%%%%%%%%%%% BODY OF PAPER %%%%%%%%%%%%%%%%%%
\section{Introduction} \label{sec:intro}
Galaxy cluster collision shocks occur when two galaxy clusters collide. These shocks are characterized by a relatively low sonic Mach number ($\MS\,\sim\,1-4$) and an Alfv{\'e}nic Mach number ($\MA$) of approximately an order of magnitude higher, leading to a plasma-$\beta$ (the ratio of thermal to magnetic energy) of approximately 100. 
The nature of these shocks has been studied extensively through cosmological hydrodynamical simulations \citep{ Miniatietal:2000,Ryuetal:2003,Pfrommeretal:2006,Kangetal:2007,Skillmanetal:2008,Hoeftetal:2008,Vazzaetal:2009, Hongetal:2014,SchaalVolker:2015,Hongetal:2015}, but whether such shocks are capable of accelerating particles to relativistic speeds, thereby contributing to the cosmic ray (CR) spectrum, remains an open question. 

Although there are indications for \emph{electron} acceleration in merging clusters \citep{vanWeerenetal:2010}, so far, studies of these shocks \citep[e.g.][]{PinzkePfrommer:2010,ZandanelAndo:2014,KangRyu:2018} have failed to find evidence of \emph{proton} acceleration. 
If CR acceleration occurred, one would expect to find evidence in the form of diffuse gamma-radiation originating from the collision between CR protons and thermal protons, followed by neutral pion decay. 
However, a study using the \emph{Fermi}-LAT instrument \citep{Ackermannetal:2016} failed to find evidence of such gamma-ray emission. 
Computational studies indicate that if galaxy cluster collision shocks produce CRs at all, the energy fraction lost to the shock in the fashion is expected to be less than one-tenth of a percent \citep{Vazzaetal:2016}. 

Further investigation of the possibility of CR acceleration in galaxy cluster shocks requires that we produce numerical models of the shock that include the necessary physical processes to follow particle acceleration. 
Although the strong shock regime has been explored extensively, using PIC, PIC-hybrid and combined PIC-MHD codes \citep[e.g.][]{RiquelmeSpitkovsky:2010,Caprioli13,CaprioliSpitkovski:2014a,Caprioli14b,Guoetal:2014a,Guoetal:2014b,Caprioli15,Baietal:2015,paper1}, as well as the Vlasov-Fokker Planck method \citep{Reville12,Reville13}, the weak shock regime has so far been largely ignored.
\citet{Haetal:2018} used particle-in-cell (PIC) simulations that showed that, depending on the input parameters, low-Mach, high-$\beta$ shocks such as these can accelerate particles, depending on the exact sonic Mach number. For those shocks with $\MS\,\simeq\,2.25$ or more, \cite{Haetal:2018} found both injection of supra-thermal particles near the shock, as well as the onset of diffusive shock acceleration (DSA), the process that allows particles to gain momentum through repeated shock crossings, also known as Fermi~I acceleration \citep[e.g.][]{Bell:1978,BlandfordOstriker:1978,Drury:1983}. 
This process requires the presence of instabilities in the local magnetic field, which can reflect the particles toward the shock. 
For shocks with a sonic Mach number of $\MS\,<\,2.25$, \citet{Haetal:2018} found a small particle injection rate (approximately an order of magnitude lower than found for the higher Mach shocks), but no DSA. 
Compared to the analytic prediction based on the Rankine-Hugoniot conditions \citep{EdmistonKennel:1984}, which placed the critical Mach number at $M_{\rm s}\approx1-1.1$, this critical Mach number (2.25) is significantly higher, which may help account for the lack of observed diffuse gamma-rays produced in low-Mach, high-$\beta$ ICM shock accelerated ions \citep{Ackermannetal:2016}.
However, the computational cost of the PIC method puts limits on the ability to follow the process over a long time and constrains the size of the spatial domain. 
Therefore, these results do not show whether the DSA process in these shocks is adequate to propel particles to relativistic speeds.

In this paper we continue the investigation of particle acceleration in weak shocks, using a different method: the particle-in-MHD-cell (PI[MHD]C) technique \citep{Baietal:2015,paper1,Mignoneetal:2018,Baietal:2019}, using the same code that was previously used in \citet{vanmarleetal:2017,casseetal:2018,vanmarleetal:2018,paper1,vanMarleetal:2019,vanMarleetal:2019b,vanmarle:2020}. 
This method allows us to take advantage of the computational efficiency of grid-based magnetohydrodynamics (MHD) while retaining the ability to follow individual, supra-thermal particles. 
We take advantage of this by running all simulations in 2-D, which will allow us to study the behaviour of the shock front and evaluate whether the effectively 1-D nature of the original simulations done by \citet{Haetal:2018} created any artefacts.
Our simulations also follow the interaction over a more extended period, which enables us to judge how the growing instabilities influence that morphology of the gas.

\subsection{Requirements for diffusive shock acceleration}
The DSA process allows a shock to accelerate particles to high relativistic speeds. However, before this process can begin, there has to be an injection of supra-thermal particles near the shock itself. 
This occurs at the shock front, with a fraction of the particles being reflected, rather than passing through the shock. The PI[MHD]C method cannot model this part of the process. 
In a PI[MHD]C simulation, shocks are modelled with MHD, and therefore regarded as discontinuities. 
The internal structure of the shock, which is far more complex  \citep[e.g.][and references therein]{Treumann:2009}, is not part of an MHD model. 
Instead, we assume that the injection process is effective, and use a parametrized injection mechanism that determines what percentage of the gas passing through the shock becomes supra-thermal. We base our injection rates on the results by \cite{Haetal:2018}, which involved the use of PIC simulations, to determine the injection rate as a function of the sonic Mach number.

Once the supra-thermal particles have been injected, further acceleration depends on their ability to generate instabilities as they travel into the upstream medium. 
If these instabilities are sufficiently strong, they will create deviations in the direction of the local magnetic field, which allows the magnetic field to reflect particles back toward the shock, starting the DSA process \citep{Sundbergetal:2016}. This process is self-sustaining and self-amplifying. The higher the momentum of the particles, the more effective they are at exciting instabilities, which, in turn, means that the efficiency of the acceleration process increases. However, the strength of the instabilities (and whether they occur at all) depends on the shock's sonic and Alfv{\'e}nic Mach numbers. 

\subsection{Layout}
This paper is structured as follows: In Sect.~\ref{sec:method} we discuss the numerical method as well as the general setup of the simulations. Next, we present a series of simulations, based on the models by \citet{Haetal:2018}, the results of which are shown in Sect.~\ref{sec-result}. Then, in Sect.~\ref{sec-longrun} we explore the long term effects of the particle acceleration process on a single, large-scale simulation. 
Finally, in Sect.~\ref{sec-conclusions}, we discuss our results and future plans. 
We also include a brief description of the equations used in our code in Appendix~\ref{sec-app1}, as well as an analytical approximation of the various time-scales involved in the perturbations in Appendix~\ref{sec-app2}.

\begin{figure*}
\centering
\mbox{
\includegraphics[width=\columnwidth]{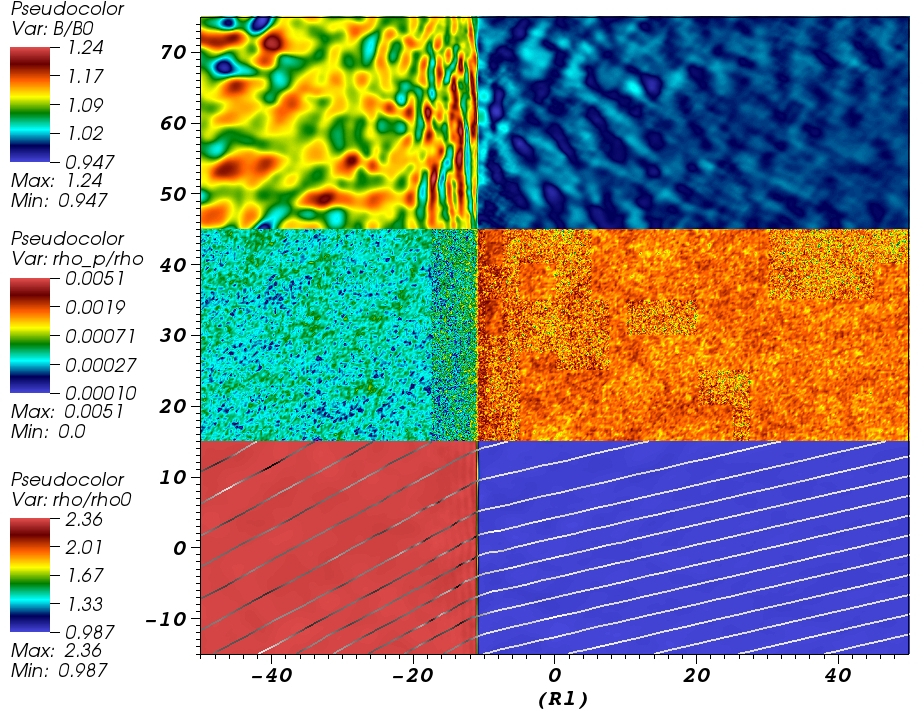}
\includegraphics[width=\columnwidth]{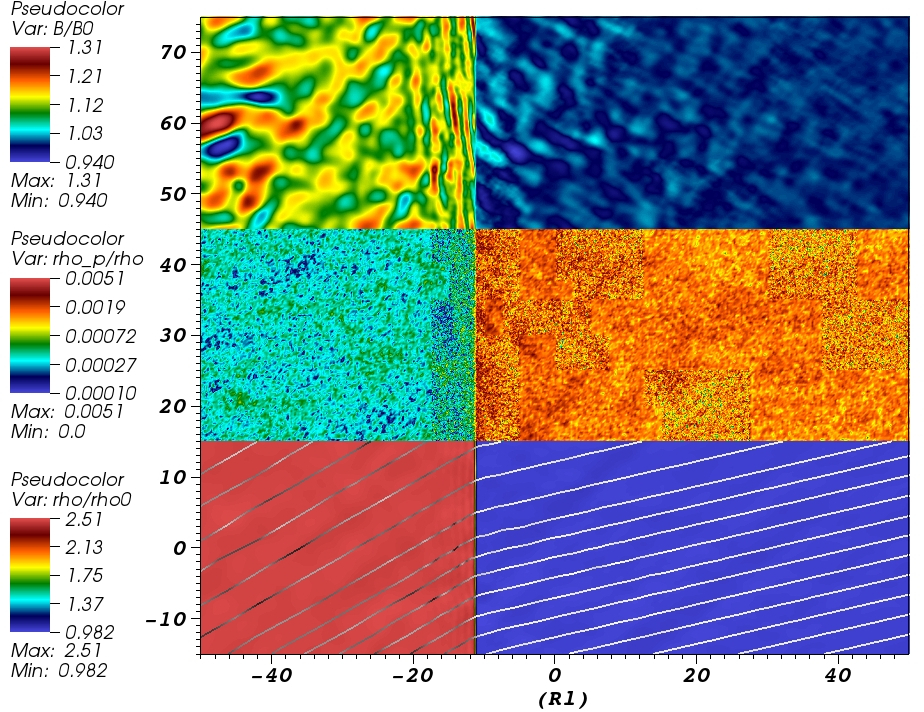}}
\mbox{
\includegraphics[width=\columnwidth]{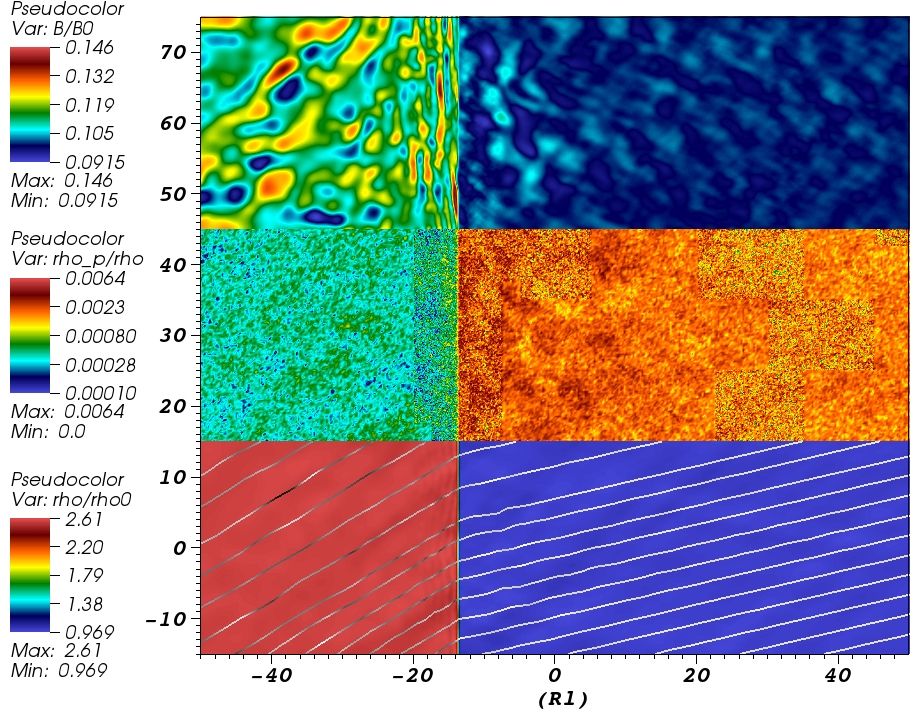}
\includegraphics[width=\columnwidth]{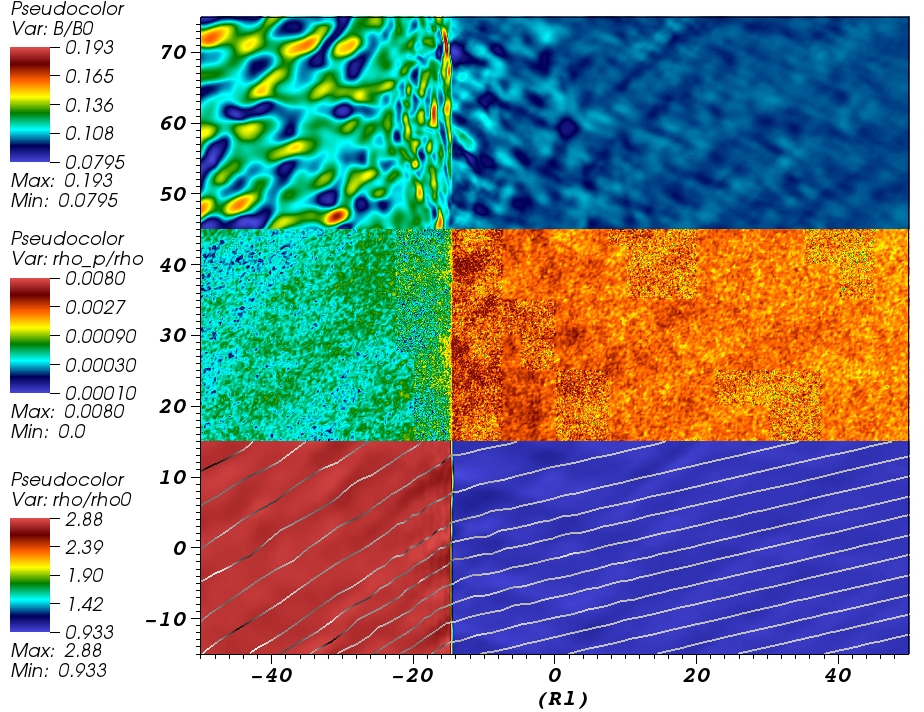}}
\caption{Morphology of the gas at the end of the simulation after $t\,=\,20\,000\,R_{\rm l}/c$, for $\MS\,=,2.0$ (top left), $2.15$ (top right) and , $2.25$ bottom left), and $2.5$ (bottom right), showing, from top to bottom the magnetic field strength relative to the undisturbed upstream magnetic field, the relative density of the supra-thermal particles relative to the thermal gas density and the thermal gas density relative to the undisturbed upstream thermal gas density. The magnetic field lines are plotted on top of the thermal gas density plot. This figure zooms in on the shock, the actual simulation box extend from $x=-90$ to $x=90$. Although all simulations show a disturbance of the magnetic field strength (top panel), none show a significant disturbance of the \emph{direction} of the magnetic field lines, which is required for particle acceleration.}
\label{fig:results1}
\end{figure*}

\begin{figure*}
\centering
\mbox{
\includegraphics[width=\columnwidth]{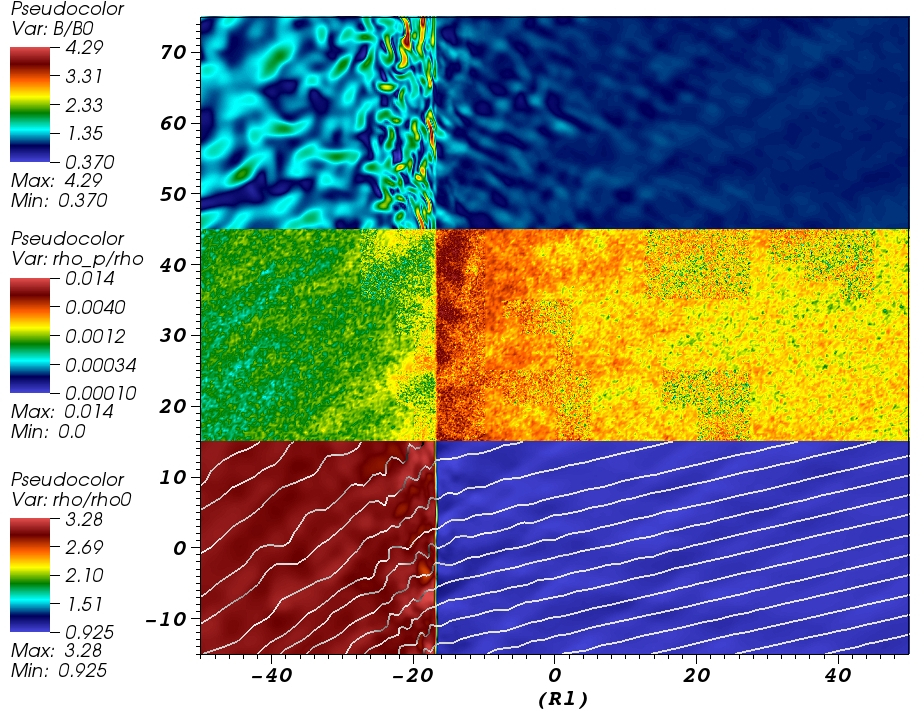}
\includegraphics[width=\columnwidth]{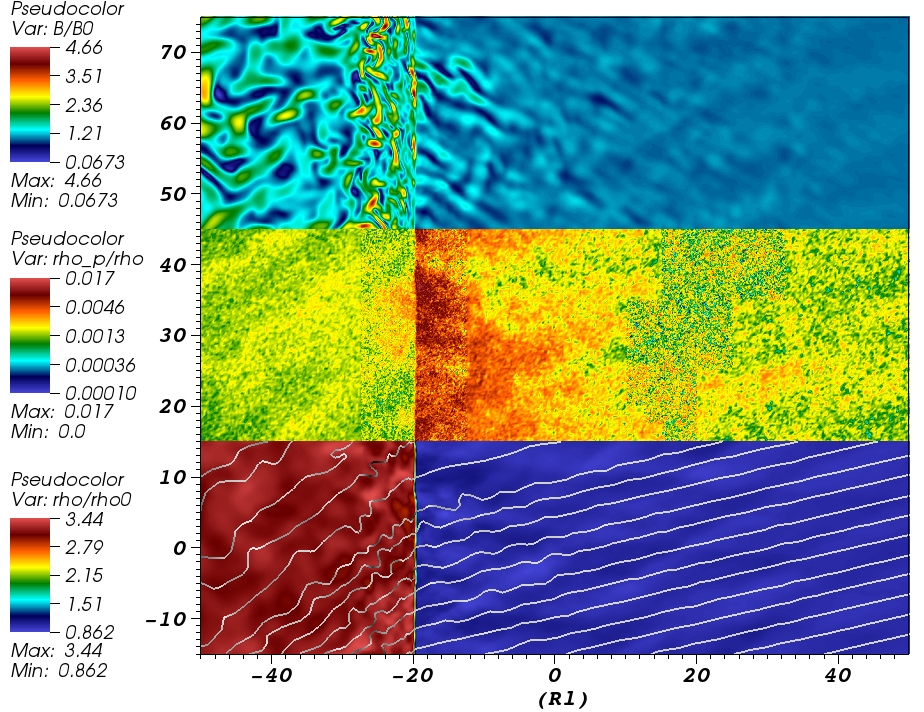}}
\mbox{
\includegraphics[width=\columnwidth]{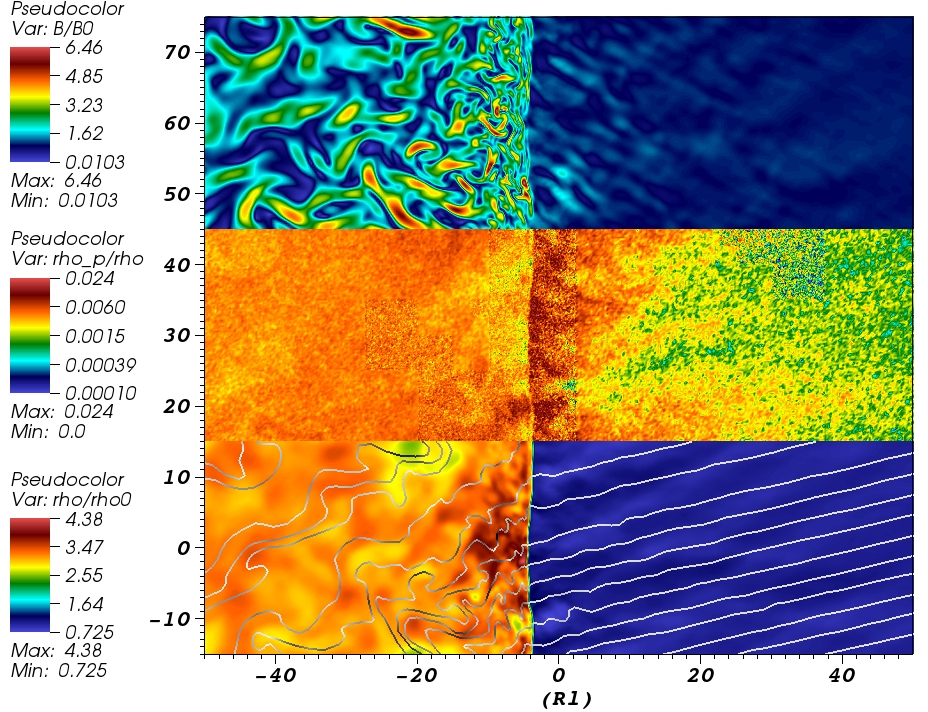}
\includegraphics[width=\columnwidth]{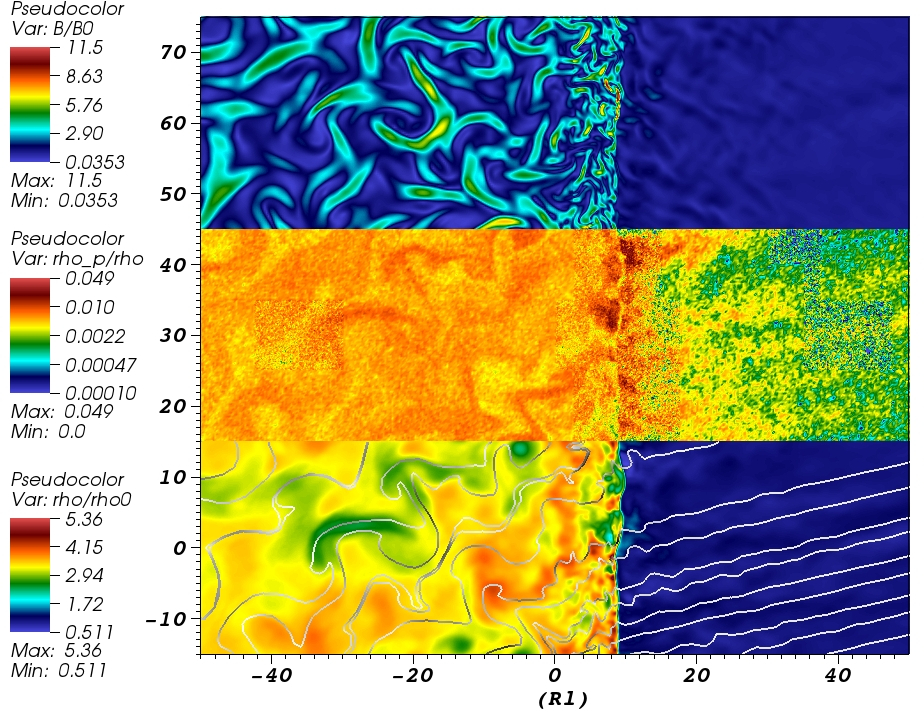}}
\caption{Similar to Fig.~\ref{fig:results1} but for simulations with for $\MS\,=,2.85$ (top left), $3.2$ (top right) and , $3.5$ (bottom left), and $4.0$ (bottom right). Unlike the simulations with lower $\MS$ these models show disturbance of the magnetic field lines, indicating that particle acceleration can take place.}
\label{fig:results2}
\end{figure*}

\section{Numerical method}
\label{sec:method}
\subsection{Code}
We use the combined PIC-MHD or Particle in MHD cells (PI[MHD]C) method described in \citet{Baietal:2015,paper1}. This method is based on the assumption that the gas can be treated as a primarily thermal gas, with a relatively small supra-thermal component. The thermal gas is treated as a fluid, using the MHD method, whereas the supra-thermal component is treated kinetically. 

The interaction between the two components is treated in a self-consistent manner. 
The force generated by the electromagnetic field of the thermal gas is part of the equation of motion of the particles, as is the opposite force on the thermal gas generated by the particles. Furthermore, the effect of a charge and current density resulting from the presence of the non-thermal particles is incorporated in the MHD-equations through a rewritten form of Ohm's law \citep{Baietal:2015},
\begin{equation}\label{Eq:Ohmslaw}
c\Efield = -\left((1-R)\vel +R\upart\right)\times\Bfield
\end{equation}
with $c$, the speed of light, $\Efield$ the electric field, $\vel$ the thermal plasma velocity, $\upart$ the average velocity of the supra-thermal particles, obtained through interpolation, $\Bfield$ the magnetic field and $R$ the supra-thermal particle charge density relative to the total charge density. 
$\upart$ and the supra-thermal particle charge density are obtained by mapping the particle distribution in space and momentum space onto the grid at the start of each time step. 
Time step control is maintained through the Courant{-}Friedrichs{-}Lewy (CFL) condition, which is applied to both the thermal gas and the particles to ensure numerical stability.

We consider the gas to be non-collisional. Therefore, there is no kinetic interaction between the particles, nor a friction force generated by the motion of the particles through the thermal gas. The only interaction between the thermal and supra-thermal components is through the electromagnetic field (the code does not include a CR-pressure term).

The main advantage of the PI[MHD]C approach is that it allows for faster calculations than either the PIC or the PIC-hybrid method, both of which require large particle populations to simulate the thermal gas, whereas the PI[MHD]C method achieves this by solving the MHD equations for the thermal gas, which is a less computationally expensive approach.

Our code is based on the {\tt MPI-AMRVAC} code \citep[e.g.][]{vanderHolstetal:2008,Keppensetal:2012}; a fully conservative, finite-volume MHD-code that solves the conservation equations of magnetohydrodynamics on an {\tt OCTREE}-based adaptive mesh \citep{ShephardGeorges:1991} and is MPI-parallel. 
To this code we have added a new set of conservation equations \citep{paper1} that incorporates the effect of supra-thermal particles on the thermal gas and a Boris-pusher \citep{BirdsallLangdon:1991} to calculate the motion of the particles as a function of the electromagnetic field. 
We have also added a constrained-transport module  based on \citet{BalsaraSpicer:1999} in order to guarantee a divergence-free magnetic field. 
As in \citet{paper1} we use a {\tt TVDLF} solver, combined with a van~Leer flux limiter. This  combination provides us with a reliable and precise numerical scheme capable of capturing the small scale features of the plasma and the magnetic field.
The conservation equations, as implemented in this code, can be found in Appendix~\ref{sec-app1}. For a full description of the numerical method as implemented in this code, please refer to \citet{paper1}, as well as the description in \citet{Baietal:2015}.

\begin{table}
 \caption{Input parameters \label{tab:input}}
 \label{tab:anysymbols}
 \begin{tabular}{lcccc}
  \hline
Model name & $\MS$ & $\MA$ & $v_0/c$ & f$_{\rm inj}$ \\
  \hline
M2     & 2.0  & 18.2 & 0.027   & 0.003 \\
M2.15  & 2.15 & 19.6 & 0.0297  & 0.003 \\ 
M2.25  & 2.25 & 20.5 & 0.0315  & 0.0035  \\
M2.5   & 2.5  & 22.9 & 0.035   & 0.0035  \\
M2.85  & 2.85 & 26   & 0.0395  & 0.00375  \\
M3.2   & 3.2  & 29.2 & 0.052   & 0.004  \\
M3.5   & 3.5  & 31.9 & 0.057   & 0.007  \\
M4     & 4.0  & 36   & 0.066   & 0.009  \\
  \hline
 \end{tabular}
\end{table}

\subsection{Simulation setup}
As input conditions, we use the same models described in \citet{Haetal:2018}, focussing on the models M2-M4 listed in Table~1 of that paper. In those simulations a beam of plasma collided with a fixed wall, forming a shock that moved upstream into the beam. These simulations, which \citet{Haetal:2018} describe as {\emph{almost 1-D}, used an extremely narrow box that covers less than a single Larmor radius in the direction perpendicular to the flow because of the high computational cost of the PIC method, which made the use of a wider box impractical. As a result, they could not resolve any variations in the thermal plasma along the perpendicular axis. \citet{Haetal:2018} defined the physics  properties of the simulations by using a fixed starting temperature for the upstream gas of $10^8\,K$ and  plasma-$\beta\,=\,100$ for all simulations. With these two quantities fixed, \citet{Haetal:2018} chose the sonic Mach number of the shock as the free parameter that made  each simulation unique. They placed upstream magnetic field in the plane of the simulation at a 13$^o$ angle with the flow.}

Because of the difference in numerical method, rather than aiming a beam of plasma at a reflecting wall and tracing the shock as it moves back into the upstream medium, we start our simulations from the analytical solution of the Rankine{-}Hugoniot conditions. This approach  allows us to use the co-moving frame of the shock as rest-frame for our simulations, reducing the required size of the simulation box along the flow direction. Like models M2-M4 in \citet{Haetal:2018}, our simulations cover the parameter space ranging from a sonic Mach number of $\MS\,=\,2-4$, with a constant plasma-$\beta\,=\,100$ and a fixed angle of $13^{\rm o}$ between the flow and the upstream magnetic field $B_0$. 
We place field in the plane of the simulation, with the third component, which is orthogonal to the plane of the simulation, initialized at zero. 
For the full set of input parameters, see Table~\ref{tab:input}, which uses the same units as in \citet{Haetal:2018}. This table lists the sonic and Alf{\'e}nic Mach numbers of the shock, the upstream velocity $v_0$ and the injection rate of supra-thermal particles. It should be noted that in this table $v_0$ is the flow-velocity in the rest-frame of the downstream medium, as used by \citet{Haetal:2018}, rather than the shock velocity, which follows from the Rankine-Hugoniot conditions. 

At the start of the simulation, we consider the gas to be purely thermal with no supra-thermal component. During the simulation, we continuously inject supra-thermal particles directly downstream of the shock, at a rate derived from the PIC simulations in \citet[][Figure~5]{Haetal:2018}. (If the shock moves during the simulation, we track this  motion to ensure that new particles continue to be injected at the shock.) 
Using these values allows us to avoid the main uncertainty in the PI[MHD]C method: the injection rate of the supra-thermal particles.
In two cases, we deviate from this pattern: the $\MS\,=\,2$ and $2.15$ simulations. \citet{Haetal:2018} found only a negligible injection rate for these particular models. 
Injecting supra-thermal particles at such a low rate would be pointless as it would merely confirm that without a 
significant fraction of supra-thermal particles the thermal gas and the shock remain in the equilibrium derived from the Rankine-Hugniot conditions. Instead, we use the power-law shown in 
\citet[][Figure~5]{Haetal:2018} to extrapolate an injection fraction of 0.003 ($0.3$\,percent of the total mass). This will allow us to determine whether any kind of DSA is possible at such a low Mach number, given the injection of a sufficient number of supra-thermal particles.

The charge-to-mass ratio of the particles reflects that of protons (we assume that all electrons are thermal) and is calculated so that a total of $1\times10^7$ particles is injected during the simulation. The total duration of each simulation equals 20\,000\,$\rl/c$, where $\rl=v_{\rm inj}/{B_0}$ the Larmor radius determined by the upstream magnetic field and the injection velocity $v_{\rm inj}$ and the upstream magnetic field $B_0$, and the injection starts at $t\,=\,1\,000\,\rl/c$, to allow the thermal gas a brief period to relax from the analytical solution to a stable numerical solution. We inject the particles with a velocity $v_{\rm inj}$ of three times the pre-shock speed of the gas, isotropically distributed within the rest-frame of the post-shock thermal gas, conform the prescription used in \citet{paper1} and \cite{vanMarleetal:2019}.
During the injection process, mass, momentum and energy are conserved by removing an equivalent amount of mass, momentum and energy density from the thermal gas in the grid cell where each particle is injected. 

For our simulations, we use a 2-D grid, spanning 180$\rl$) along the x-axis (parallel to the flow) by 30$\rl$ along the z-axis (perpendicular to the flow). At the coarsest level, our grid has one grid cell per Larmor-radius, and we allow three additional levels of refinement. Each level doubles the effective resolution. Therefore, at maximum resolution, we have eight grid cells per Larmor radius. 
For comparison with PIC or di-hybrid simulations, we need to keep in mind that the particle injection velocity is three times the upstream bulk velocity in the rest-frame of the shock. Therefore $\rl$, as defined in this paper, is three times the Larmor radius of the upstream thermal gas. 
Within five Larmor radii of the shock, we enforce the highest level of refinement at all times to ensure that both the shock itself and the particle motion near the shock remain fully resolved. 
Vector quantities (velocity, electromagnetic field) have three components, making our models effectively 2.5-D. When calculating the charge and current density resulting from the particle distribution, we map the charge and current of each particle onto the four surrounding cell-centres.

\begin{figure}
\centering
\mbox{
\includegraphics[width=\columnwidth]{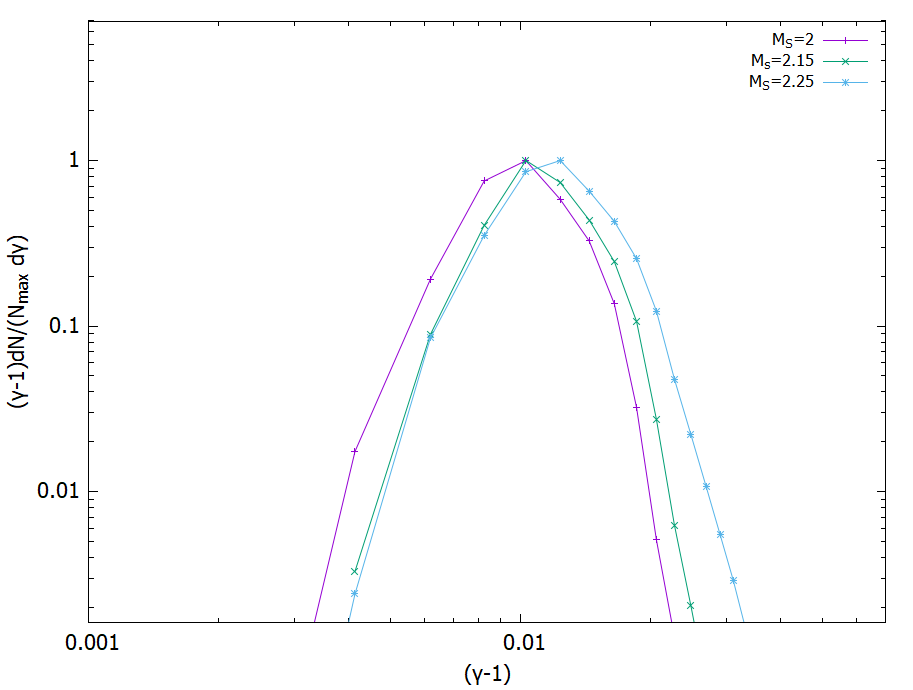}}
\caption{Normalized spectral energy distribution, as a function of the Lorentz factor ($\gamma$), for the total particle population of the simulations with sonic Mach numbers 2-2.25 at $t\,=\,20\,000\,R_{\rm l}/c$. None of these SEDs show any evidence for DSA.}
\label{fig:parased1}
\end{figure}

\begin{figure}
\centering
\mbox{
\includegraphics[width=\columnwidth]{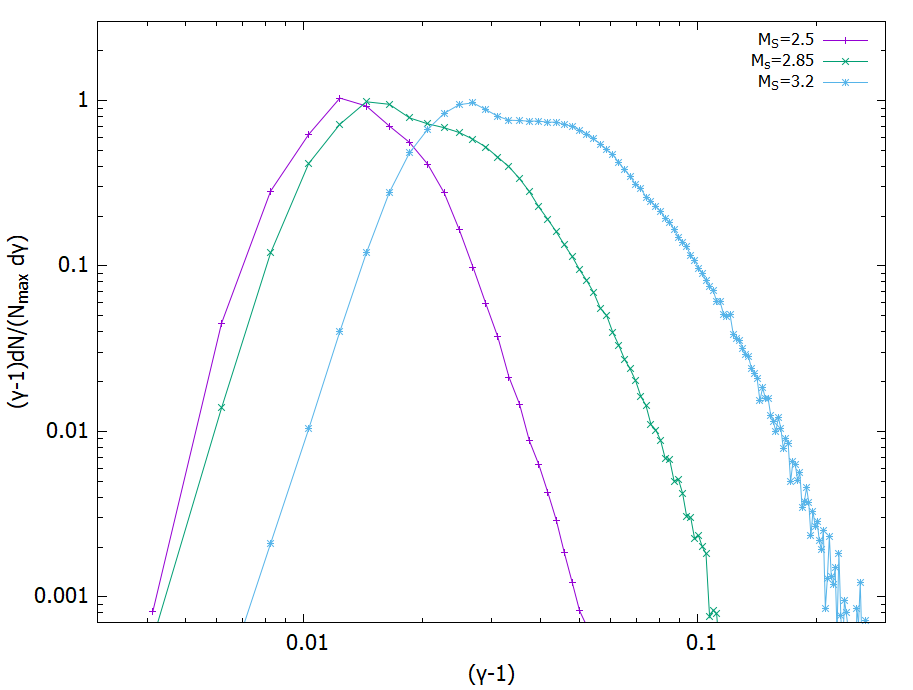}}
\caption{Similar to Fig.~\ref{fig:parased1} but for the simulations with sonic Mach numbers 2.5-3.2. These simulations show evidence of DSA, but only for part of the particle population.}
\label{fig:parased2}
\end{figure}

\begin{figure}
\centering
\mbox{
\includegraphics[width=\columnwidth]{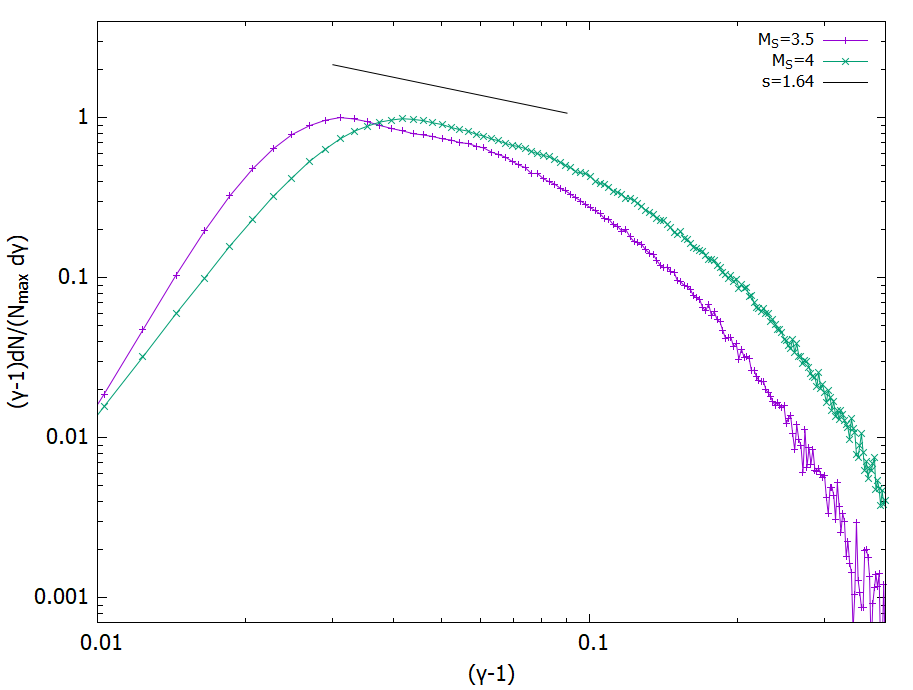}}
\caption{Similar to Figs.~\ref{fig:parased1}-\ref{fig:parased2} but for the simulations with sonic Mach numbers 3.5 and 4. These simulations show clear evidence of DSA but only for part of the particle population. The slopes are approaching the expected power law index.  For comparison we have added the analytically predicted slope for the $M_S\,=\,4$ shock ($s\,=\,1.64$).}
\label{fig:parased3}
\end{figure}

\section{Parameter space}
\subsection{2-D Morphology of the gas}
\label{sec-result}
The response of the thermal gas and the magnetic field to the injection of the supra-thermal particles varies according to the sonic Mach number, as shown in Figs.~\ref{fig:results1}-\ref{fig:results2}, shows a clear trend. Each panel shows the magnetic field strength relative to the undisturbed upstream magnetic field strength ($B_{\rm 0}$) (top), the supra-thermal gas density ($\rho_{\rm p}$) relative to the thermal gas density (middle) and thermal gas density relative to the undisturbed upstream thermal gas density ($\rho_{\rm 0}$) (bottom) as well as the magnetic field lines. Although the simulations are run in the shock rest-frame (determined from the analytical solution), the shocks tend to move over time. 
The injection process removes a small but noticeable fraction of the thermal gas pressure in the downstream medium, which would lead the shock to start moving in the downstream direction. 
Depending on the strength of the instabilities, this trend can be reversed if the downstream magnetic field amplification becomes sufficiently strong to counteract this effect by increasing local magnetic pressure.
The low-Mach simulations ($\MS\,\leq\,2.5$) show a disturbance of the magnetic field strength, but not of the magnetic field lines, even though we have artificially enhanced the injection rate for the $\MS\,=\,2-2.15$ shocks.
As a result, even though the magnetic field is influenced by the presence of supra-thermal particles, no DSA can be expected to occur. In contrast, the (relatively) high-Mach simulations ($\MS\,\geq\,3.2$) show disturbance of both the field lines and the field strength, indicating that DSA can take place. In between these two extremes lies a regime that shows some distortion of the field lines, but only locally, with some field lines showing twists, while others do not. 
Even in those cases where the perturbations are relatively strong, they tend to be weaker than  perturbations found for comparable models of high-Mach, low-$\beta$ shocks \citep{paper1}. 
The magnetic field amplification tends to be of the order of $2-4$, not exceeding one order of magnitude even for the $\MS\,=\,4.0$ model. For comparison, the $\MS\,=\,30.0$, $\beta\,=\,1$ shocks in \citet{paper1} showed amplification of 15-25, depending on the angle between the flow and the magnetic field.

This result can be compared qualitatively to the results found by \citet{Haetal:2018}, which determined that the instabilities in the lower Mach simulations are resonant-streaming instabilities \citep{Bell:1978}, whereas the higher Mach models display both resonant and non-resonant streaming \citep{Bell04} characteristics. 
This behaviour is consistent with the results of \citet{Caprioli14b}, which determined that the transition between resonant and non-resonant streaming instabilities lies around an Alf{\'e}nic Mach number of $\MA\,\simeq\,30$. 
The critical Mach number, below which no significant instabilities occur, seems to lie somewhat higher than what was found by \citet{Haetal:2018}, who put it at ($\MS\,\simeq\,2.25$) This can be explained by the nature of the injection mechanism, which, in our case, is isotropic in the post-shock medium, whereas, in reality, the particle injection velocity and direction is a more complicated issue that depends on the exact conditions of the shock. 
Furthermore, our models show a transition region ($2.5\,\geq\,\MS\,\geq\,3.2$), rather than a single critical Mach number. 
This is the result of having a fully 2-D model, rather than one that is effectively 1-D In a 1-D model, the magnetic field is either perturbed or not, whereas a 2-D model allows for variations along the plane of the shock. As will be seen in Sect.~\ref{sec-parased}, these variations also influence  the particle spectral energy distribution (SED). 

The instabilities lack the filamentary structure that is characteristic of the high-Mach, low-$\beta$ simulations \citep[e.g.][]{paper1,vanMarleetal:2019}. 
Such filaments require a relatively stiff magnetic field that can resist the forces that are acting upon it. 
In a high-$\beta$ plasma, the magnetic field is powerless to resist either the distorting force exerted by the particles (assuming a sufficiently high injection rate) or the thermal pressure. 
As a result, a local instability will tend to grow in all directions, distorting the magnetic field, rather than conforming to the direction of the field lines. 

How the perturbations in the upstream medium depend on the sonic Mach number can be explained by a comparison of three different time-scales: The perturbing time-scale, which depends on the force acting on the magnetic field, determines how quick the instabilities grow. The sonic and magnetic time-scales determine how quickly the thermal gas can respond. 
As long as the sonic and magnetic time-scales are shorter than the perturbing time-scale, the instabilities cannot grow. On the other hand, if the perturbing time-scale is shorter than both the sonic and magnetic time-scales, the instabilities will continue to grow because the thermal gas cannot compensate in time. 
In between these two extremes lies a regime where the perturbing time-scale is shorter than the magnetic time-scale, which allows the variations in magnetic field strength to grow, but longer than the sonic time-scale, which means that any perturbation in the direction of the field lines is counteracted by the thermal pressure. 
(N.B. This only applies to the high-$\beta$ regime. In gas with a plasma-$\beta\,<\,1$, the situation would be reversed.) An analytical approximation of the various time-scales is presented in Appendix~\ref{sec-app2} but should be treated with caution because it depends on many factors, which tend to vary over time. 
This time-scale dependence explains the behaviour of the high-$\beta$ simulations. 
In the case of the $\beta\,=\,1$ simulations, which have been more common in the past, the thermal gas either shows a clearly perturbed magnetic field, or it shows no instabilities at all because the 'in-between' regime does not exist. The larger the gap between the sonic and Alf{\'e}nic speeds, the more opportunities there are for a 'partial instability' to occur where only the field strength varies, whereas the field direction remains constant.

For most simulations, the plane of the shock remains almost undisturbed, in contrast to the high-Mach shocks shown in \citet[e.g.][]{Baietal:2015,paper1,vanMarleetal:2019}, where ram-pressure variations in the upstream medium, resulting from the filamentary structure of the instabilities, caused the corrugation and eventually the complete distortion of the shock front. For low-Mach shocks, the variation in the total pressure that the shock experiences is much smaller and both time- and space-dependent. As a result, the shocks show little to no corrugation. 

\subsection{Spectral distribution of supra-thermal particles}
\label{sec-parased}
The momentum distribution for particles accelerated through diffusive shock acceleration is given by 
\begin{equation}
\label{eq:mom}
f(p)~\propto~p^{-q}
\end{equation}
with $q\,=\,3r_{\rm c}/(r_{\rm c}-1)$ and $r_{\rm c}$ the compression ratio of the shock \citep{BlandfordOstriker:1978}. 
For non-relativistic particles where the particle energy $E_{\rm i}$ scales with the momentum as $E_{\rm i}\,\propto\,p_{\rm i}^2/m_{\rm i}$. 
This results in a spectral energy distribution that can be represented by 
\begin{equation}
\label{eq:gammamin1}
\frac{\partial N}{\partial\gamma}~\propto~(\gamma-1)^{-s},
\end{equation}
with $s\,=\,(q-1)/2$ and $\gamma$ the Lorentz factor. However, should the particles attain relativistic speeds the energy/momentum relationship changes until, at high relativistic speeds, it becomes $E_{\rm i}\,\propto\,p_{\rm i}$. 

The SEDs for the particles, shown in Figs.~\ref{fig:parased1}-\ref{fig:parased3} in our simulations can be classified into three distinct groups. Like \citet{Haetal:2018}, we find no evidence for DSA for the lowest sonic Mach numbers ($\MS\,=\,2.0-2.25$, Fig.~\ref{fig:parased1}). 
The particles show no sign of significant acceleration, as was already predicted from the morphology of the gas and the magnetic field. Neither is there any sign that the particles adhere to a power-law spectrum, which would appear as a straight line in a log-log plot. Instead, the particle energies have spread out around the injection speed, indicating that although some particles have been accelerated, others have actually been decelerated. Simulations of high-Mach shocks \citep{paper1,vanmarle:2020} also show that a fraction of the injected particles loses decelerate, triggering instabilities in the thermal plasma at the expense of their own energy. However, in high-Mach shocks, once the instabilities have been triggered, subsequent interactions between the magnetic field accelerate particles through the DSA process. 
For the $\MS\,=\,2.0-2.25$ simulations shown here, no significant acceleration occurs because the instabilities dissipate almost immediately. 
The end result is a particle energy distribution that starts to resemble that of a thermal plasma. 
However, because the particles do not collide directly, they cannot actually become thermalized. 
The behaviour at low-Mach numbers is particularly meaningful in light of the artificially enhanced injection rate that we have adopted for the models with $\MS\,=\,2.0-2.15$. 
Despite the high injection rate of supra-thermal particles, there is no evidence of DSA. 
Clearly, under these conditions, the shock is incapable of accelerating CRs.

At the other end of the scale ($\MS\,=3.5-4$, Fig.~\ref{fig:parased3}) the particles SEDs show clear evidence for DSA. Both SEDs show evidence that the energy distribution is starting to follow  a straight line in the log-log plot, indicating adherence to a power-law with a fixed index between $(\gamma-1)\,=\,0.03-0.06$ for the $\MS\,=\,3.5$ shock and  $0.04-0.08$ for the $\MS\,=\,4$ shock. For the latter, the expected slope, following from the compression ratio  would be $s\,\sim\,1.64$ according to Eq.~\ref{eq:gammamin1}. We have included an indicator of that expected slope in the plot for comparison.

For the intermediate Mach numbers (Fig.~\ref{fig:parased2}, the SEDs show a combination of these two extremes.  
Although there is evidence for DSA, in the form of an extended tail at high energies, there is also a peak around the injection energy, clearly indicating that not all particles are participating in the acceleration process. 
That some particles can avoid participating in the DSA process is partially the result of the behaviour of the magnetic streamlines described in Sect.~\ref{sec-result}. 
Depending on the injection location, particles may either encounter instabilities on its journey upstream or not, giving the particles a chance to escape from the system without being accelerated. Adherence to a power-law index is barely visible, except in the $\MS\,=\,3.2$ case, and even there only for a very short interval, although this is partially a result of the limited size of the simulation box. As the particles gain momentum, they will travel further away from the shock before being reflected. If they reach the upper or lower x-boundary before being reflected, they escape from the system. Although these simulations give us a good indication whether DSA is possible at all, in order to estimate its efficiency as well as the maximum possible energy, we will need to increase the size of the simulation box. (See Sect.~\ref{sec-longrun}).

\begin{figure*}
\centering
\mbox{
\includegraphics[width=\columnwidth]{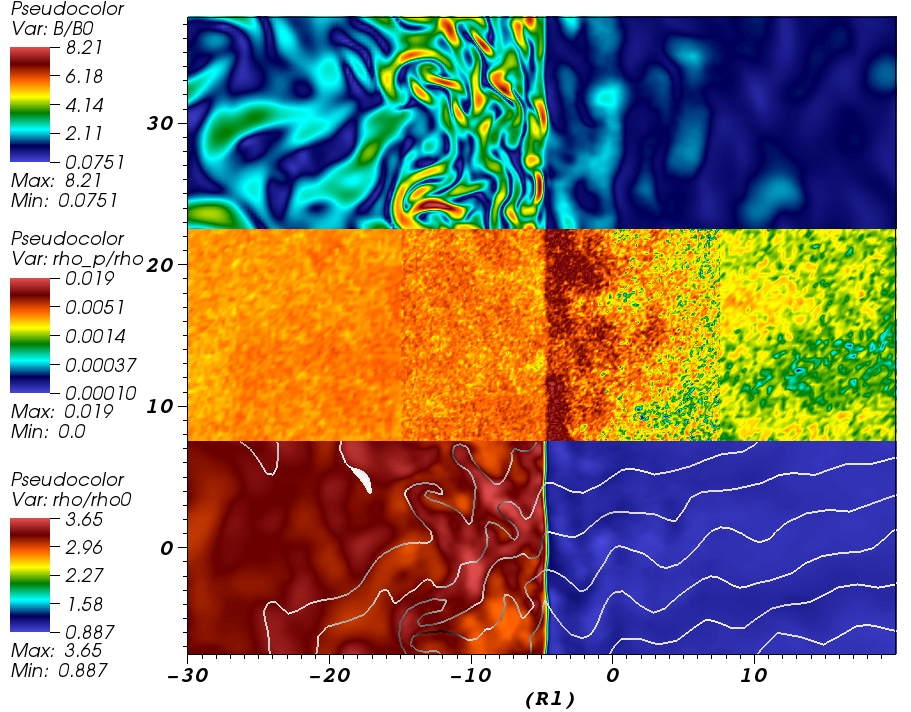}
\includegraphics[width=\columnwidth]{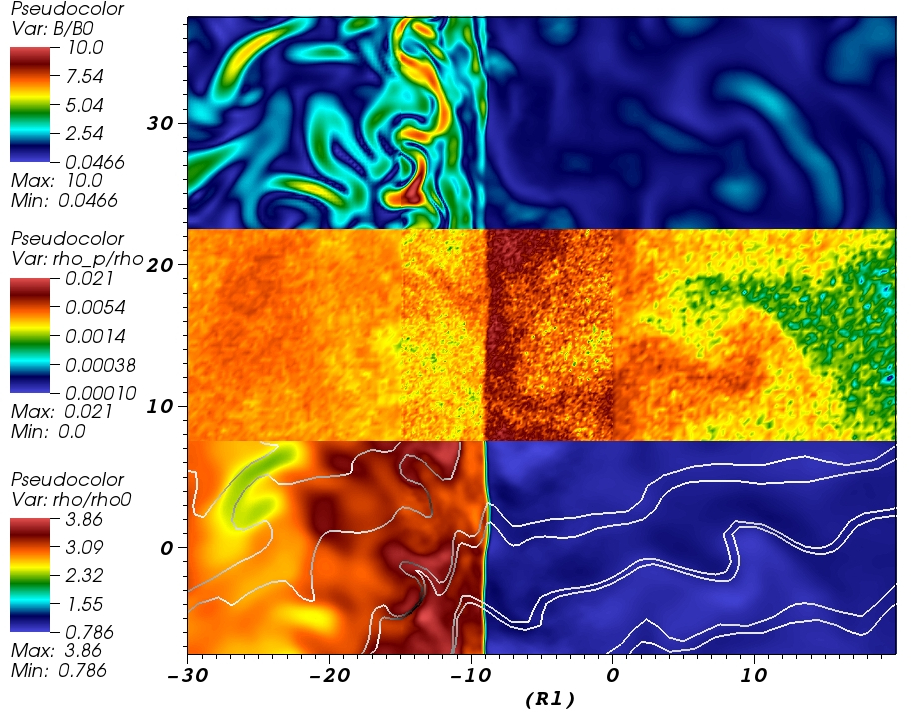}}
\mbox{
\includegraphics[width=\columnwidth]{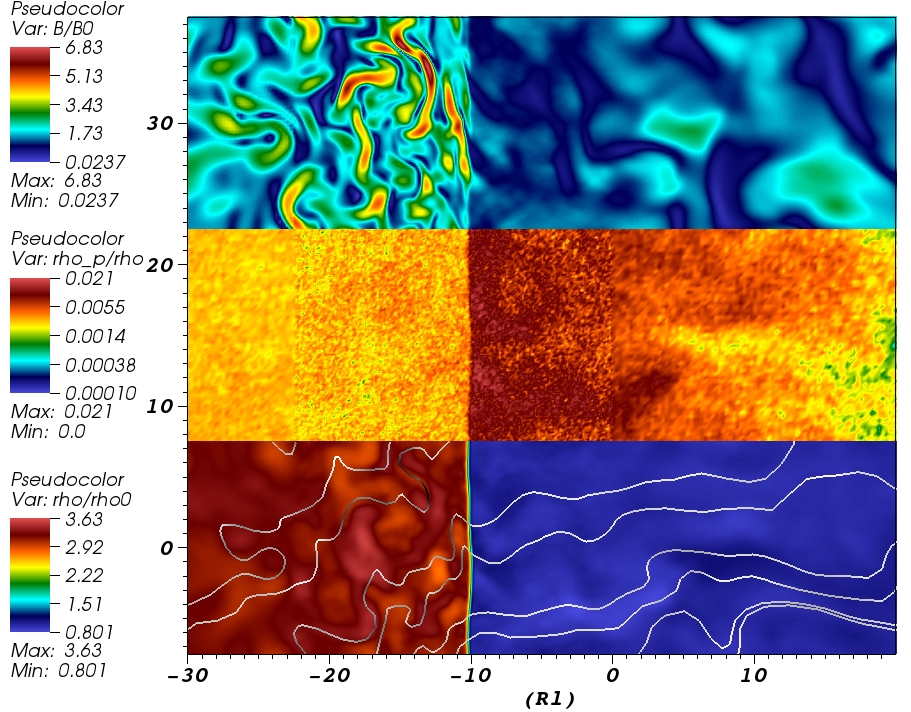}
\includegraphics[width=\columnwidth]{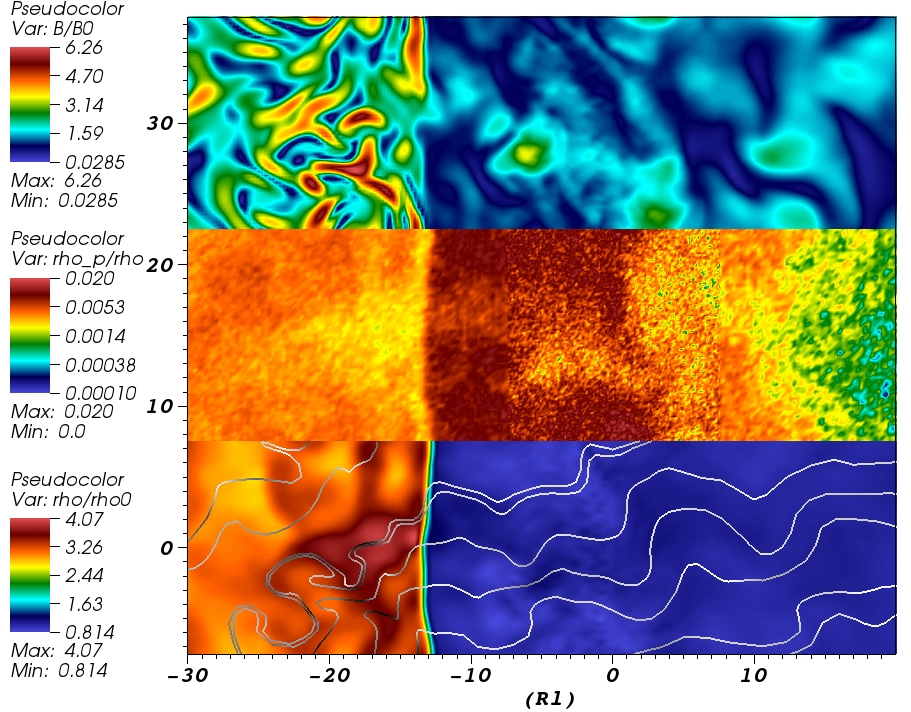}}
\caption{Similar to Figs.~\ref{fig:results1}-\ref{fig:results2}, but for the $\MS\,=\,3.2$ model with an extended simulation box, showing the area near the shock. These figures show the area near the shock at $t\,=\,5\,000$ (top left), 10\,000 (top right) 
15\,000 (bottom left) and 20\,000 $R_{\rm l}/c$ (bottom right). The large distortion of the magnetic field is clearly visible. }
\label{fig:results3}
\end{figure*}

\begin{figure}
\centering
\mbox{
\includegraphics[width=\columnwidth]{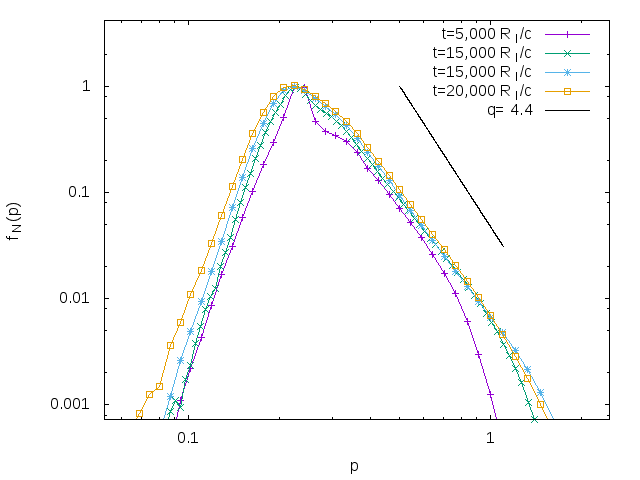}}
\caption{Total particle SED at the same moments in time as \ref{fig:results3} for the  $\MS\,=\,3.2$ model with an extended simulation box. Over time, the SED matches the expected power-law index and shows that DSA can accelerate particles to relativistic speeds. 
The $q\,=\,4.4$ line demonstrates the power-law slope expected for a Mach 3.2 shock.}
\label{fig:longboxSED}
\end{figure}

\begin{figure}
\centering
\mbox{
\includegraphics[width=\columnwidth]{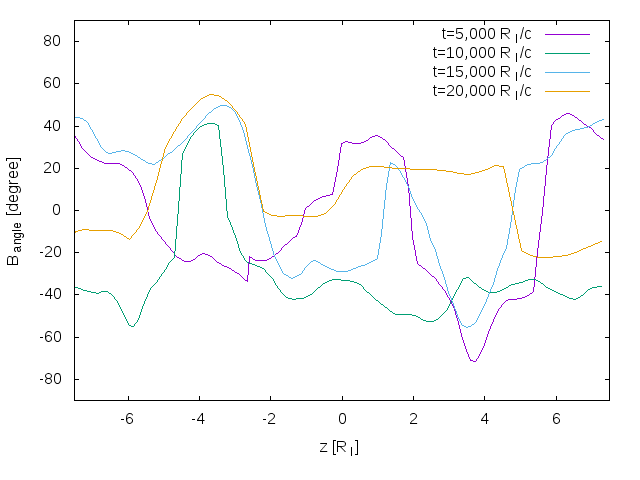}}
\caption{Angle between the flow and the magnetic field at the shock as a function of the location along the z-axis at the same moments in time as in Fig.~\ref{fig:results3}. The shock, which started out as quasi-parallel, occasionally becomes semi-perpendicular as a result of the instabilities in the upstream medium.}
\label{fig:Bangle}
\end{figure}

\begin{figure}
\centering
\mbox{
\includegraphics[width=\columnwidth]{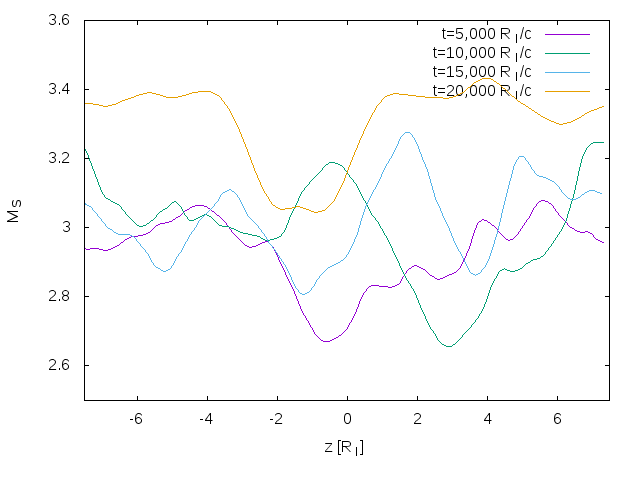}}
\caption{Sonic mach number at the shock as a function of the location along the z-axis at the same moments in time as in Figs.~\ref{fig:results3} and \ref{fig:Bangle}. Whereas the shock started at $\MS\,=\,3.2$, it varies over time between approximately 2.7 and 3.4.}
\label{fig:machnumber}
\end{figure}

\section{The large box model}
\label{sec-longrun}
In order to explore the long term effects of the instabilities as well as to determine the maximum particle energy that can be achieved in this fashion, we repeat the simulation with a sonic Mach number of 3.2 with a far more extended box, which has a size of $2400\,\times\,15\rl$, compared to the $180\,\times\,30\rl$ box of the previous simulations. (We have reduced the size of the bow along the z-axis by half in order to reduce computation time.)
For this simulation, we use a minimum resolution of $4800\,\times\,30$ grid points and allow two additional levels of refinement in order to achieve the same effective resolution as in the models described in Sect.~\ref{sec-result}. The result, in a series of snapshots, is presented in Fig~\ref{fig:results3}, which shows the morphology of the gas and the magnetic field near the shock at $t\,=\,5\,000$ ), 10\,000, 15\,000, and 20\,000 $R_{\rm l}/c$.  
As for the smaller box model, this simulation shows that at Mach 3.2, the upstream magnetic field becomes distorted, facilitating the DSA. As can be seen in Fig.~\ref{fig:results3}, the box size along the z-axis is still large enough that it does not interfere with the maximum wavelength of the instabilities and allows for variation along the plane of the shock. 

\subsection{The particle SED}
\label{sec-parasedlong}
Figure~\ref{fig:longboxSED} (right panel) shows the particle SED the total particle populations at the same moments in time as Fig.~\ref{fig:results3}. 
Because the particles are entering the relativistic regime, where the relationship between energy and momentum changes, we show the SED as a function of normalized momentum $p=\gamma v$. Using the Lorentz factor would no longer produce a straight line in the log-log plot. For a $\MS\,=\,3.2$ shock, the power-law index can be expected to be approximate $q\,=\,4.4$, according to Eq.~\ref{eq:mom}. We have added a slope indicator in the plot for comparison. The slope matches reasonably well, though not perfectly. 
However, we should keep in mind that the power-law index depends on the compression ratio of the shock. 
For the $\MS\,=\,3.2$ shock, the compression ratio is expected to be a factor 3.1, based on the Rankine-Hugoniot conditions. However, as will be seen in Sect.~\ref{sec-distort}, the sonic Mach number is not a constant in time owing to the instabilities in the upstream medium.

The extreme length of the simulation box ensures that the particles can continue to be part of the acceleration process, rather than escaping from the simulation, allowing them to reach much higher velocities. 
The SED shows adherence to a power-law up to approximately eight times the injection momentum, reaching a Lorentz factor of approximately 1.8, which, for protons, is the equivalent of a total energy of $E_{\rm tot}\,=\,\gamma mc^2\,=\,1.6$\,GeV.
Such protons, if they were to interact with thermal protons, would be able to produce gamma-radiation. Beyond this point, the SED drops off, and, from the time evolution, it is clear that running the simulation longer will not change the energy at which the DSA ceases to be effective.

\subsection{Distortion of the upstream medium}
\label{sec-distort}
As the instabilities grow in size, they start to change the nature of the shock. 
The angle at which the magnetic field enters the shock varies considerably, both in time and space, as shown in Fig.~\ref{fig:Bangle}, which shows the angle between the flow and the magnetic field directly ahead of the shock at different moments in time. \citet{Haetal:2018} found that the angle of the magnetic field with the shock does not greatly influence the particle injection rate, as long as the shock qualifies as quasi-parallel. 
Similarly, \citet{Fangetal:2019} found comparable ion SEDs for shocks varying between 15$^{\rm o}$ and 45$^{\rm o}$.
However, this relies on the magnetic field remaining quasi-parallel. As can be seen in Fig.~\ref{fig:Bangle}, which shows the angle between the flow and the magnetic field ahead of the shock, this can no longer be assumed because the magnetic field becomes quasi-perpendicular at certain times, depending on the position along the shock front. How this will influence the injection rate is a complex question. 
\citet{Haetal:2018} found that a quasi-perpendicular shock could give particles a limited acceleration through the shock-drift acceleration (SDA) process, but found no evidence for DSA. This result coincides with the results found by \citet{CaprioliSpitkovski:2014a} for higher Mach shocks, whereas \citet{paper1} found that DSA could occur in quasi-perpendicular shocks, assuming a sufficiently high injection rate. 
However, none of the above results genuinely apply to a situation in which only part of the shock is quasi-perpendicular and then only at certain times.
A further effect of the increased instabilities in the upstream gas is a change in the sonic Mach number of the shock, shown in Fig.~\ref{fig:machnumber} for the same moments in time as in Fig.~\ref{fig:Bangle}. As the magnetic field becomes distorted, the flow in the thermal gas starts to compress the loops in the magnetic field as they approach the shock. The compression of the magnetic field, in turn, increases the local magnetic field strength, which, combined with the magnetic tension as the curvature of the magnetic field lines increases allows the magnetic field to resist the pressure effectively. This causes variations in the local gas temperature and flow-velocity, changing the Mach number. 
The influence of the sonic Mach number on the injection rate is shown quite clearly by \cite{Haetal:2018}, though we should keep in mind that this was for simulations with a fixed plasma-$\beta$, something that does not apply to our simulations as a result of the changes in the upstream morphology.

\section{Conclusions}
\label{sec-conclusions}
We have investigated the behaviour of low-Mach, high-$\beta$ shocks and their ability to accelerate particles through the DSA process, using the PI[MHD]C method. 
For small scale simulations, our results bear a close similarity to those obtained by \citet{Haetal:2018} with PIC simulations, allowing for the inherent uncertainty caused by the fact that we have to assume a particle injection rate. Rather than finding a 'critical Mach number', at which the shocks go from not accelerating particles at all to clear evidence of DSA, we find that there is a transitional zone between approximately $\MS\,\approx\,2.5-3.2$. For shocks with Mach-numbers below this zone, no acceleration takes place, even when stimulated with an artificially high injection rate, while above this zone the shocks show clear evidence of DSA. 
Within the transitional zone, some particle acceleration will occur, but the SEDs show deviation from the expected power-law index. 
It is likely that within this region the efficiency of DSA will depend strongly on input conditions, which, in turn, means that some galaxy cluster shocks in this parameter space will be able to accelerate CRs, whereas others will not, depending on local circumstances.

Over longer time-scales, the growing instabilities in the upstream medium observed in simulations with sonic Mach number of >3 deviate from the results by \citet{Haetal:2018}. This deviation takes the form of varying Mach numbers at the shock, as well as a severely distorted magnetic field. Once the simulation reaches this point, the fixed injection rate is no longer a valid assumption. 
In the long run, this effect is likely to cause a considerable change in the efficiency of CR production. Even those shocks that have a sufficiently high Mach number will end up exciting such severe instabilities that the Mach number starts to vary. Only if this causes the injection rate to be sufficiently reduced, which is certainly possible if the Mach number drops, the instabilities will fade, which will eventually allow the shock to return to its original strength, at which point the process will repeat itself. 
Furthermore, the power-law index of the particle SED depends directly on the compression ratio  of the shock. 
Owing to the instabilities, this will become increasingly variable over time, limiting our ability to quantify the energy loss through cosmic rays, which is unlikely to be a constant factor. Further investigation, in particular regarding the injection efficiency as a function of Mach-number, plasma-$\beta$, and magnetic field angle, is required before such an analysis becomes possible.

These simulations demonstrate that large-scale, multi-D simulations are required in order to thoroughly investigate the behaviour of shocks in the presence of supra-thermal particles. While 1-D simulations can tell us much about the structure of the shock itself as well as the injection process, the long term evolution of the instabilities that enable the acceleration process is a fundamentally multi-D problem on scales that exceed the Larmor radius of the injected particles. This requires us to use simulation boxes that can capture the large-scale variations in the thermal plasma perpendicular to the flow and extend sufficiently to allow for repeated particle acceleration events.

Future developments will have to include a scheme that adjusts the injection rate as a function of shock conditions. However, this is not an easy task because the shock conditions do not only vary over time but depend on the location along the shock front as well. 
Furthermore, the velocities of the thermal gas in this kind of shock is approaching the point where a relativistic treatment of the thermal gas becomes necessary. A new version of the PI[MHD]C code, which allows for the combination of a relativistic thermal gas with non-thermal particles, is currently undergoing testing. 

Whether there is merit to repeating these simulations in 3-D is an open question. \citet{vanMarleetal:2019} showed for a high-Mach model that the morphology of the thermal gas and the magnetic field did not change significantly between 2-D and 3-D models, but that the SED of the 3-D model showed a marked decrease in acceleration efficiency. However, that was for a shock that became severely distorted over time, something that is not the case for the low-Mach models. As it is, the computational cost of a 3-D model, particularly a 3-D model large enough to resolve the instabilities and follow the acceleration of particles to high velocities, remains prohibitive.

\section*{Acknowledgements}
This work was supported by the National Research Foundation (NRF) of Korea through grant 2016R1A5A1013277 and by Basic Science Research Program through the National Research Foundation of Korea (NRF) funded by the  Ministry of Education, Science and Technology (2018R1D1A1B07044060). This work was supported by the National Research Foundation (NRF) of Korea through grant 2016R1A5A1013277 and by Basic Science Research Program through the National Research Foundation of Korea (NRF) funded by the  Ministry of Education, Science and Technology (2018R1D1A1B07044060). This work is supported by the ANR-19-CE31-0014 GAMALO project. 

The author wishes to thank Prof. D. Ryu, Prof. H. Kang and J.-H. Ha for their valuable comments and discussions.

The author thanks the anonymous reviewer for their many helpful comments

\section*{Data availability}
Data available on request.

%%%%%%%%%%%%%%%%%%%%%%%%%%%%%%%%%%%%%%%%%%%%%%%%%%



%\bibliographystyle{mnras}


\bibliographystyle{yahapj}
\bibliography{vanmarle_biblio.bib}

%%%%%%%%%%%%%%%%%%%%%%%%%%%%%%%%%%%%%%%%%%%%%%%%%%

%%%%%%%%%%%%%%%%% APPENDICES %%%%%%%%%%%%%%%%%%%%%

\appendix
\section{Numerical scheme}
\label{sec-app1}
Our code solves the MHD conservation equations for mass, momentum and energy, with additional terms that incorporate the interaction with the non-thermal particles. The continuity equation remains unchanged, 
\begin{equation}
\frac{\partial \rho}{\partial t} ~+~ \nabla \cdot (\rho \vel)~=~0,
\label{eq:mass}
\end{equation}
where $\rho$ and $\vel$ stand for the mass density and velocity of the thermal plasma. The momentum equation has to be adjusted to incorporate the additional force exerted by the particles.
\begin{equation}
\frac{\partial \rho\vel}{\partial t} + \nabla\cdot\left(\rho\vel\otimes\vel-\frac{\Bfield\otimes\Bfield}{4\pi}+P_{\mathrm tot}\unit\right)~=~-\Fpart,
\label{eq:momentum}
\end{equation}
with $\Bfield$ the magnetic field and $P_\mathrm{tot}=P+B^2/8\pi$ the total pressure. The energy equations becomes
\begin{equation}
\frac{\partial e}{\partial t}+\nabla\cdot\biggl( (e + P_{\rm tot})\vel+(\Efield-{\mathbf E}_0)\times\frac{\Bfield}{4\pi}\biggr) 
~=~ -\upart \cdot \Fpart
\label{eq:energy}
\end{equation}
with $e$ the total energy density of the thermal plasma. The new terms on the right hand side of Eq.(\ref{eq:momentum},\ref{eq:energy}) are dependent on an averaged supra-thermal particle velocity
$\upart$ as well as the opposite of the force density that the thermal plasma exerts on the supra-thermal particles $\Fpart$, which can be determined as
\begin{equation}
\label{eq:lorentz}
\Fpart~=~(1-R) \biggl( \npart e {\mathbf E}_0 + \frac{\Jpart}{c} \times \Bfield \biggl), 
\end{equation}
with $\npart$ and $\Jpart$ the supra-thermal particle density and the current generated by the supra-thermal particles.  
The term $\mathbf E_0$ equals the electric field generated by the thermal plasma alone, whereas  $\Efield$ is the total electric field produced by the thermal and non-thermal components combined.
We close the equations with an updated form of the Maxwell-Ampere equation,
\begin{equation}
\frac{\partial\Bfield}{\partial t}~=~c\nabla \times \Efield
\end{equation}

Meanwhile the motion of the particles is controlled by the particle momentum equation:
\begin{equation}
\frac{\partial \mathbf{p}_{\alpha ,j}}{\partial t}~=~q_j\biggl(\Efield+ \frac{{\mathbf u}_{\alpha,j}}{c}\times\Bfield \biggr)
\end{equation}
with ${\mathbf p}_{\alpha ,j}$, $q_j$ and ${\mathbf u}_{\alpha ,j}$ the momentum, charge and velocity of the particle with index $j$.

For a full derivation of these equations we refer to \citet{paper1} as well as \cite{Baietal:2015}.

\section{The time-scales of instabilities}
\label{sec-app2}
Whether or not the presence of supra-thermal particles can cause significant instabilities in the upstream medium depends on the balance between three different time-scales: The time-scale of the force causing the disturbance and the two time-scales at which the plasma responds (the sonic and Alfv{\'e}nic time-scales). 
To do a first-order analysis, we can determine the sonic response as
\begin{equation}
\frac{\partial x}{\partial t}~=~c_{\rm s}
\end{equation}
and the magnetic response as
\begin{equation}
\frac{\partial x}{\partial t}~=~c_{\rm A}
\end{equation}
with $c_{\rm s}$ and $c_{\rm A}$ the sound speed and Alfv{\'e}n speed respectively.
The time-scale of the distorting force can be approximated by the square root of the acceleration divided by the length scale, 
\begin{equation}
\frac{\partial^2 x}{\partial t^2} ~=~\frac{F_{\rm l}}{\rho}
\end{equation}
with $F_{\rm l}$ the Lorentz force and $\rho$ the thermal gas density. 
Integrating over a characteristic length scale (for which we choose the gyro radius $r_l$) gives us the sonic time-scale,
\begin{equation}
\tau_{\rm s}~=~\frac{r_{l}}{c_{\rm s}}
\end{equation}
and the magnetic time-scale
\begin{equation}
\tau_{\rm B}~=~\frac{r_{l}}{c_{\rm A}}
\end{equation}
Finally, the perturbing time-scale can be found as,
\begin{equation}
\begin{aligned}
\tau_{F}~&\propto&~\sqrt{\frac{r_{l}\rho}{(1-R)\rho_{part} \mathbf{E}_0~+~n_{part}|\mathbf{v}_{part}\times \mathbf{B}|}}
\end{aligned}
\end{equation}
by substituting the Lorentz force exerted by the particles on the local magnetic field (see Eq.~\ref{eq:lorentz} for the Lorentz force in the PI[MHD]C model) and normalizing the speed of light as well as the proton-mass and charge to 1. For simplicity we assume that the magnetic field is parallel to the flow, therefore the background electric field $E_0$ is zero. 
This simplifies the characteristic time-scale to, 
\begin{equation}
\begin{aligned}
\tau_{F}~&\sim&~\sqrt{\frac{r_{l}\rho}{\rho_{part}|\upart\times \mathbf{B}|}} \\    
            &=&~\sqrt{\frac{r_{l}^2\rho}{\rho_{part}|\upart\times\mathbf{v}_{fluid}|}}, 
\end{aligned}
\end{equation}
because $B=\gamma \frac{|\upart|}{r_l}$  assuming (again normalizing proton charge and mass to one). The Lorentz factor $\gamma$ can be set to one as well because we consider the thermal gas to be non-relativistic. 
So, the time-scale is ultimately defined as, 
\begin{equation}
\begin{aligned}
\tau_{F}~&\sim&~\sqrt{\frac{r_{l}\rho}{\rho_{part}|\upart\times \mathbf{B}|}} \\    
            &=&~r_{l}\sqrt{\frac{\rho}{\rho_{part}}} \frac{1}{\sqrt{|\upart| v}}, 
\end{aligned}
\end{equation}
As long as this time-scale is shorter than the other two, the fluid cannot effectively respond and the magnetic field lines will start to twist. If it is longer than both, nothing will happen at all. 
If it is in between these two, things get complicated. In the typical scenario where the sound speed exceeds the Alfv{\'e}n speed, this would mean that although the magnetic field strength starts to vary, the field lines remain aligned because the sound waves smooth out any variation. 
Obviously, this is a very rough approximation, but it can be used for qualitative estimates of the behaviour of the flow.

%%%%%%%%%%%%%%%%%%%%%%%%%%%%%%%%%%%%%%%%%%%%%%%%%%

% Don't change these lines
\bsp	% typesetting comment
\label{lastpage}
\end{document}